\documentclass[pre,aps,floats,superscriptaddress,floatfix,twocolumn]{revtex4}
\usepackage{amssymb,amsmath}
\usepackage{amsmath,amssymb}
\usepackage{graphicx}
\usepackage{psfrag}
\usepackage{color}
\usepackage{dcolumn}
\usepackage{bm}
\usepackage[normalem]{ulem}

\def\beq{\begin{equation}}
\def\eeq{\end{equation}}
\def\bea{\begin{eqnarray}}
\def\eea{\end{eqnarray}}
\begin{document}

\title{Universal properties of the Kardar-Parisi-Zhang equation with quenched columnar  disorders}
\author{Astik Haldar}\email{astik.haldar@gmail.com}
\affiliation{Condensed Matter Physics Division, Saha Institute of
Nuclear Physics, HBNI, Calcutta 700064, India}

\begin{abstract}
 Inspired by the recent results on totally asymmetric simple exclusion processes on a periodic lattice with short-ranged quenched hopping rates [A. Haldar, A. Basu, Phys Rev Research {\bf 2}, 043073 (2020)],  we  study the universal scaling properties of the Kardar-Parisi-Zhang (KPZ) equation with short-ranged quenched columnar disorder in general $d$-dimensions. We show that there are generic propagating modes in the system that have their origin in the quenched disorder and make the system anisotropic. We argue that the presence of the propagating modes actually make the effects of the quenched disorder irrelevant, making the universal long wavelength scaling property belong to the well-known KPZ universality class. On the other hand, when these waves vanish in a special limit of the model, new universality class emerges with dimension $d=4$ as the lower critical dimension, above which the system is speculated to admit a disorder-induced roughening transition to a perturbatively inaccessible rough phase.
\end{abstract}

\maketitle

\section{Introduction}

Quenched disorders are known to significantly affect the universal scaling properties of condensed matter systems and statistical models in equilibrium. For example, the universal critical properties of classical $N$-component spin models near its second order phase transitions can be affected by the quenched disorder, leading to a new universality class~\cite{quench-on,quench-on1}. On the other hand, quenched random fields can destroy the ordered phase altogether~\cite{rand-field}.  Scaling properties of interface growth models are also found to be affected by quenched disorders \cite{kim96, vicsek93, barabasi96, barabasi94}.  Understanding of how quenched disorders affect universal properties in nonequilibrium systems is far less understood than their equilibrium counterparts, and remains a challenging problem. Lack of any general framework to study the statistical properties of nonequilibrium systems has often prompted physicists to construct and study simple conceptual models that are easily analytically tractable, such that questions of basic principles can be answered within well-controlled approximations. 

Recently, the universal scaling properties of the density fluctuations in a totally asymmetric simple exclusion process (TASEP) in a ring geometry with quenched disordered hopping rates has been studied by mapping the problem into a one-dimensional ($1d$) Kardar-Parisi-Zhang equation with columnar disorder~\cite{astik-prr}. It has been shown that away from the half-filling, the quenched disorder is irrelevant, and the universal long wavelength properties of the system is identical to that of the $1d$ KPZ equation without any disorder. On the other hand, close to half-filling, quenched disorder is relevant, a new universality class different from the $1d$ KPZ universality emerges, leading to the scaling exponents taking values different from their $1d$ KPZ counterpart.  These results qualitatively agree with the simulations performed \cite{mustansir1, mustansir2, stinchcombe} for driven lattice gases in presence of quenched disorder .

In this article we generalise the $1d$ KPZ equation with short-ranged quenched columnar disorder~\cite{astik-prr} to $d$-dimensions, and study its universal scaling properties in the long wavelength limit. Our main results in the article are as follows. (i) The generalised $d$-dimensional equation does not remain invariant under Galilean transformation in presence of the disorder, and there are generic underdamped propagating waves in the system. The latter makes the system necessarily ansiotropic. (ii) Surprisingly, the presence of the waves makes the disorder {\em irrelevant} (in a renormalisation group or RG sense). As a result, the  Galilean invariance and isotropy are restored in the long wavelength limit; the universal scaling property naturally belongs to the $d$-dimensional KPZ universality class. (iii) For specific choices of the model parameters, the waves vanish and  the system is isotropic. In this case the disorder is relevant (in a RG sense), and the Galilean invariance remains broken in the long wavelength limit. Consequently, the universal scaling exponents belong to a new universality class different from the pure KPZ equation. We argue that $d=4$ is the {\em lower critical dimension} of the model in this case, such that above $d=4$, the model can undergo a roughening transition to a (perturbatively inaccessible) rough phase. We use a one-loop dynamic RG calculational framework for our work.

The rest of this article is organised in the following manner. We set up the hydrodynamic equations for height of growing surfaces with quenched columnar disorder in Sec. \ref{Model}. The KPZ universality has been discussed in Sec. \ref{KPZ univ}. We discuss the scaling behaviour  in linear theory in Sec.~\ref{lin}, and  then calculate the nonlinear effects on the scaling exponents by using RG framework for this quenched disordered model in rest of Sec.~\ref{scaling}. We summarise our results in Sec. \ref{summary}. Some technical details are provided in Appendix for the interested readers.

\section{KPZ equation with columnar quenched disorder}\label{Model}

We  begin by recalling the $1d$ hydrodynamic equation, derived in Ref.~\cite{astik-prr} for density fluctuations in a periodic TASEP with quenched disordered hopping rates in the thermodynamic limit. The equation of motion (EoM) for  density fluctuation $\phi(x,t)$ reads~\cite{astik-prr}
\begin{eqnarray}
 \frac{\partial \phi(x,t)}{\partial t}&=& \nu \frac{\partial^2 \phi(x,t)}{\partial x^2} +  \lambda_1 \frac{\partial \phi(x,t)}{\partial x}+ \lambda_m \frac{\partial \delta m (x)}{\partial x}  \nonumber \\ &&    - \frac{\lambda}{2} \frac{\partial  \phi^2(x,t)}{\partial x} + \lambda_2 \frac{\partial}{\partial x}[\delta m(x) \phi(x,t)]  \nonumber\\&&  - \lambda_3 \frac{\partial}{\partial x}[\delta m (x) \phi^2(x,t)] + \frac{\partial f(x,t)}{\partial x}.\label{phieq}
\end{eqnarray}
 Here, $\nu$ is a diffusion coefficient,  $\lambda$, $\lambda_m$, $\lambda_1$, $\lambda_2$, $\lambda_3$ are different model parameters. In particular, $\lambda_1$, $\lambda_2$ are both proportional to $(2\rho_0-1)$ where $\rho_0$ is the mean particle density. Hence, $\lambda_1$, $\lambda_2$ vanish in the half-filled limit ($\rho_0=1/2$) whereas those have non zero value for away from the half-filled limit~\cite{astik-prr}.  Furthermore, $m(x)$ is the random quenched disorder that has its origin in the quenched disorder hopping rates in the underlying lattice-gas model. As explained in Ref.~\cite{astik-prr}, density fluctuation $\phi(x,t)$ is driven by {\em two} conserved stochastic variables: quenched noise $\partial_x\delta m (x)$ and annealed noise $\partial_x f(x,t)$.
   We write $m(x)= m_0 + \delta m(x)$, where $m_0=\overline{m(x)}$ is the mean value of $m(x)$, and $\delta m(x)$ is local fluctuation of $m(x)$ about $m_0$. We assume $\delta m(x)<m_0$ to ensure $m(x)$ remains positive everywhere; we choose $\delta m(x)$ to be short-ranged Gaussian distributed with a variance
\begin{equation}
 \overline{ \delta m(x)\delta m(0)} = 2\tilde D\delta (x);\label{mvari1}
\end{equation}
$\tilde D$ is the strength of the quenched noise, and is definite positive.

Stochastic function $f(x,t)$ is an annealed noise which is assumed to be zero-mean and Gaussian-distributed with  a variance
\begin{equation}
 \langle f(x,t) f(0,0)\rangle = 2D\delta(x)\delta (t).\label{noise-vari1}
\end{equation}
Here, $D>0$ is the noise strength, the analogue of temperature in nonequilibrium systems.  Here $\langle...\rangle$ implies the averages over time-dependent noise distribution and an overline implies the averages over quenched disorder distribution.

The variances of $f(x,t)$ and $\delta m(x)$ in the Fourier space are
\begin{subequations}
 \begin{align}
 &\langle f(k_1,\omega_1) f(k_2,\omega_2)\rangle = 2D\delta (k_1+k_2)\delta(\omega_1 +\omega_2),\label{noise-variF}\\
 & \overline {\delta m(k_1,\omega_1)\delta m(k_2,\omega_2)}=2\tilde D\delta (k_1+k_2)\delta(\omega_1 +\omega_2)\delta(\omega_1).\label{mvariF}
 \end{align}
\end{subequations}

Let us introduce a height variable via the transformation $\phi(x,t)=\partial_x h$. This transforms (\ref{phieq}) to
\begin{eqnarray}
 \frac{\partial h(x,t)}{\partial t}&=& \nu \partial_{xx} h(x,t) +  \lambda_1 \partial_x h(x,t)+ \lambda_m \delta m (x)  \nonumber \\ &&    - \frac{\lambda}{2} [\partial_x  h(x,t)]^2 + \lambda_2 [\delta m(x) \partial_x h(x,t)]  \nonumber\\&&  - \lambda_3 [\delta m (x) (\partial_x h(x,t))^2] +  f(x,t).\label{heq}
\end{eqnarray}
 Here $h(x,t)$,  a single-valued function, is the height of the $1d$ surface with respect to an arbitrary baseline. Equation~(\ref{heq}) gives the dynamics of a growing $1d$ surface with quenched columnar disorder. { When $\lambda_1=0=\lambda_2$, Eq.~(\ref{heq}) is invariant under inversion of space, i.e., under $x\rightarrow -x$, whereas for $\lambda_1\neq 0,\, \lambda_2\neq 0$, it is not due to the presence of the propagating waves. These correspond, respectively, to near the half-filled and away from the half-filled limits for density fluctuation in Eq.~(\ref{phieq})}. 
 
 { We now generalise the problem to a $d$-dimensional growing surface in presence of columnar quenched disorder. This can be done in several ways. We generalise in a way to ensure that the $d$-dimensional generalised equation has a definite symmetry, and reduces to the $1d$ case in a simple and unambiguous way. } Noticing that there are two terms (with coefficients $\lambda_1$, $\lambda_2$) containing single spatial derivative on $h$ in (\ref{heq}) which imply the absence of the space inversion symmetry in EoM, we generalise the EoM of $h({\bf x},t)$ to $d$-dimensions by assuming breakdown of the inversion symmetry along one particular dimension, while the system remains inversion-symmetric along all other $(d-1)$ directions. We use $\parallel$  as subscript  to denote the special direction in which the inversion symmetry is absent.  The generalised EoM for $h({\bf x},t)$ ($\bf x$  now refers as position on $d$-dimensional surface) has the generic form
\begin{eqnarray}
 \frac{\partial h}{\partial t}&=& \nu \nabla^2 h +  \lambda_1 \partial_\parallel h+ \lambda_m \delta m  - \frac{\lambda}{2} [ \boldsymbol\nabla  h]^2   \nonumber \\ && + \lambda_2 [\delta m \partial_\parallel h] - \lambda_3 [\delta m ( \boldsymbol\nabla h)^2] +  f.\label{heqd}
\end{eqnarray}
It is clear that Eq.~(\ref{heqd}) has a lower symmetry than the $d$-dimensional KPZ equation. It is rotationally invariant in the $(d-1)$-dimensional subspace, that excludes the ``$\parallel$''-direction, unlike the $d$-dimensional KPZ equation that is rotationally invariant in the full $d$-dimensional space. It is instructive to write down the ``Burgers'' version of Eq.~(\ref{heqd}) for a ``quenched disorder Burgers velocity'' ${\bf v}$ by using the transformation ${\bf v}={\boldsymbol\nabla} h$:

\begin{eqnarray}
 \frac{\partial {\bf v}}{\partial t}&=& \nu \nabla^2 {\bf v} +  \lambda_1 \boldsymbol\nabla v_\parallel+ \lambda_m \boldsymbol\nabla  \delta m  - \frac{\lambda}{2} \boldsymbol\nabla [ v]^2   \nonumber \\ && + \lambda_2  \boldsymbol\nabla[\delta m  v_\parallel] - \lambda_3  \boldsymbol\nabla[\delta m ( v)^2] +   \boldsymbol\nabla f.\label{veqd}
\end{eqnarray}

The variances of the noises in $d$-dimensional space are just the direct generalisations, respectively, of  (\ref{mvari1}) and (\ref{noise-vari1}):
\begin{subequations}
 \begin{align}
 &\overline{ \delta m({\bf x})\delta m(0)} = 2\tilde D\delta^d ({\bf x}).\label{mvari}\\
  &\langle f({\bf x},t) f(0,0)\rangle = 2D\delta^d({\bf x})\delta (t).\label{noise-vari}
  \end{align}
\end{subequations}

 Due to the presence of quenched disorders, Eq.~(\ref{heqd}) does not remain invariant under the Galilean transformation ${\bf x}\mapsto {\bf x}+{\bf v}_0t, \,t\mapsto t,\, h({\bf x},t\mapsto h({\bf x}+{\bf v}_0t)+{\bf v}_0\cdot {\bf x})$.  In Eq.~(\ref{heqd})  the term with linear derivative $\lambda_1 \partial_\parallel h$ implies the existence of underdamped propagating modes. The surface would be anisotropic due to presence of the propagating modes with picking up a specific direction, that without any loss of generality is assumed to be along the $x$-direction, i.e., { the ``parallel direction'' $x_\parallel$ is identified with the $x$-direction}. In the absence of any quenched disorder $(\delta m({\bf x})=0)$,  Eq.~(\ref{eqh2d})  reduces to the well-known KPZ equation.

 In what follows below, it is convenient to write $h({\bf x},t)=h_1({\bf x})+h_2({\bf x},t)$~\cite{astik-prr}. Time-independent function $h_1({\bf x})$ then satisfies
\begin{eqnarray}
 && -\nu_\psi\nabla^2h_1 - \lambda_{1\psi} \partial_\parallel h_1 - \lambda_m \delta m + \frac{\lambda_{\psi}}{2} ( \boldsymbol\nabla h_1)^2 \nonumber \\&&- \lambda_{2\psi}[\delta m (\partial_\parallel h_1)] + \lambda_{3\psi} [\delta m ( \boldsymbol\nabla h_1)^2] = 0.\label{eqh1d}
\end{eqnarray}
Time-dependent function $h_2({\bf x},t)$  satisfies
\begin{eqnarray}
 \frac{\partial h_2}{\partial t} &=& \nu_{\rho} \nabla^2 h_2 + \lambda_{1\rho} \partial_\parallel h_2 - \frac{\lambda_{\rho}}{2} ( \boldsymbol\nabla h_2)^2 + \lambda_{2\rho}[\delta m (\partial_\parallel h_2)]   \nonumber \\&-&  \lambda_{\rho\psi}( \boldsymbol\nabla h_1)\cdot ( \boldsymbol\nabla h_2) - \lambda_{3\rho} [\delta m ( \boldsymbol\nabla h_2)^2]  \nonumber\\ && - \lambda_{3\rho\psi}[\delta m  \boldsymbol\nabla h_1 \cdot  \boldsymbol\nabla h_2] + f .\label{eqh2d}
\end{eqnarray}
Here, we use different sets of model parameters associated $h_1$ and $h_2$ to allow for different scalings of $h_1$ and $h_2$ in long wavelength limit. The parameters $\nu_\psi, \nu_\rho$ are come from $\nu$; $\lambda_\psi$, $\lambda_{\rho}$, $\lambda_{\rho\psi}$ are from $\lambda$;  $\lambda_{1\psi}$, $\lambda_{1\rho}$ are from $\lambda_1$; $\lambda_{2\psi}$, $\lambda_{2\rho}$ are from $\lambda_2$ and $\lambda_{3\psi} $, $\lambda_{3\rho}$, $\lambda_{3\rho\psi}$ are from $\lambda_3$. 

Equations~(\ref{eqh1d}) and (\ref{eqh2d}) obviously generalise the corresponding $1d$ equations in Ref.~\cite{astik-prr}. Field $h_1({\bf x})$ represents the frozen height of a surface which is entirely ``driven'' by the quenched disorder $\delta m({\bf x})$, whereas $h_2$ represents the time-dependent height of a growing surface, driven by the time-dependent additive noise $f$. { Clearly, the dynamics of the height field $h_2({\bf x},t)$ in  Eq.~(\ref{eqh2d}) is affected by the frozen height field $h_1({\bf x},t)$ that satisfies Eq.~(\ref{eqh1d}), and hence by the quenched disorder}. 
 
 We define two cases depending on the choices of parameters. {\em Case-I:} Here, $\lambda_1\neq 0$, $\lambda_2\neq 0$ hence $\lambda_{1\psi},\, \lambda_{1\rho},\, \lambda_{2\psi},\, \lambda_{2\rho} $ are nonzero,  leading to the presence of propagating modes. {\em Case-II:} here, $\lambda_1= 0$, $\lambda_2= 0$ mean $\lambda_{1\psi}=\lambda_{1\rho}= \lambda_{2\psi}=\lambda_{2\rho}=0 $,  propagating modes vanish and the full $d$-dimensional isotropic property of surface is restored. These two cases are  corresponding to the away from half filled limit and near to the half-filled limit of the periodic TASEP \cite{astik-prr} respectively.
 
\section{Kardar-Parisi-Zhang universality class}\label{KPZ univ}
We first briefly review the KPZ universality before analysing our disordered model. The KPZ equation is given by~\cite{kpz}
\begin{equation}
 \frac{\partial h}{\partial t} = \nu \nabla^2 h -\frac{\lambda}{2} ({\boldsymbol \nabla} h)^2 + f.\label{kpz}
\end{equation}
Here, $h({\bf x},t)$ is height of $d$-dimensional surface at any instant time $t$ and driven by Gaussian white noise $f$ with satisfying (\ref{noise-vari}). This model equation admits Galilean invariance due to transformation of the inertial frame. 
Correlation function of the height fluctuations follows the scaling in the long wavelength limit:
\begin{equation}
 \langle [h({\bf x},t) - h(0,0)]^2\rangle = |{\bf x}|^{2\chi}\varTheta(|{\bf  x}|^{z}/t).
\end{equation}
Here, $\chi$ and $z$ are roughness and dynamic exponents for $h$; $\varTheta$ is a dimensionless scaling function of its argument. 
One could find the scaling behaviour of $h({\bf x},t)$ in the long time limit by defining a dimensionless coupling constant $g=\frac{\lambda^2D}{\nu^3}$. The RG flow  equation for $g$ in the one-loop perturbative theory satisfies
 \begin{equation}
  \frac{d g}{dl}=  g \left[2-d+g\frac{4d-6}{d}\right].\label{kpz-flow}
 \end{equation}
 Notice from (\ref{kpz-flow}) that $g$ diverges at $d=3/2$; furthermore, between $3/2<d<2$ $g=0$, is the only physically acceptable solution. This is believed to be an artefact of the one-loop perturbation theory~\cite{Frey-two-loop}.
The Galilean invariance of the KPZ equation leads to  $\chi+z=2$, an exact relation between the scaling exponents. At $1d$,  $z=3/2$ and $\chi=1/2$ are exactly known analytically due to the fluctuation-dissipation-theorem in the model. Coupling  $g$ grows under rescaling and is marginally relevant at $d=2$, the lower critical dimension of the KPZ equation. 
The KPZ equation admits a roughening transition (smooth-to-rough phase transition) for dimensions higher than $2d$; the rough phase  is inaccessible in the perturbative theory.

\section{Universal properties and scaling behaviour}\label{scaling}

 The autocorrelation functions of $h_1({\bf x})$ and $h_2({\bf x},t)$ are characterised by the universal scaling exponents. The autocorrelation functions are given by
\begin{eqnarray}
  &&C_{1}({\bf x})\equiv\overline{ [h_1({\bf x})-h_1(0)]^2}= |{ x}_\parallel|^{2\chi_1}f_1\left(\frac{|{\bf x}|^{\mu_1}_\perp}{x_\parallel}\right),\\
  && C_{2}({\bf x},t)\equiv \overline{\langle [h_2({\bf x},t) - h_2(0,0)]^2\rangle} \nonumber\\&&~~~~~~~~~~~= |x_\parallel|^{2\chi_2} f_2(|{ x_\parallel}|^z/t,\frac{|{\bf x}|^{\mu_2}_\perp}{x_\parallel}).
\end{eqnarray}
Here, $f_1$ and $f_2$ are two dimensionless scaling functions of their arguments; ${\bf x}=(x_\parallel,{\bf x}_\perp)$ with $x_\parallel$ being the direction of propagation of the traveling waves { and ${\bf x}_\perp$ being $d-1$ transverse directions to the propagating direction of waves}. The scaling functions represent the generic anisotropy in the presence of the propagating waves. Scaling exponents 
$\chi_1$ and $\chi_2$ are roughness exponents for $h_1$ and $h_2$ respectively and $z$ is dynamic exponent of $h_2$; exponents $\mu_1,\,\mu_2$ are the anisotropy exponents with $\mu_2=1$ in the pure $d$-dimensional KPZ equation (since it is fully isotropic).

\subsection{Linear theory}\label{lin}
  The equal-time autocorrelation functions of $h_1({\bf x})$ and $h_2({\bf x},t)$ can be written down exactly in the linear theory using the linear terms in (\ref{eqh1d}) and (\ref{eqh2d}). These in the Fourier space   for {\em Case-I} are
\begin{subequations}
 \begin{align}
   &\langle |h_1({\bf k})|^2 \rangle= \frac{2\tilde{D}\lambda^2_m}{\lambda_{1\psi}^2 k_\parallel^2+ \nu_\psi^2 k_\perp^4},\label{h1-cor}\\
 &\langle |h_2({\bf k},t)|^2 \rangle= \frac{D}{\nu_\rho k^2},\label{h2-cor}
 \end{align}
\end{subequations}
where in (\ref{h1-cor}) we have ignored $k_\parallel^4$ in comparison with $k_\parallel^2$ in the long wavelength limit.
 Naturally, the exponents $\chi_1=(2-d)/2$ and $\chi_2=(2-d)/2$ in the linear theory are known exactly.{ Furthermore, we identify that $\mu_1=2$ and $\mu_2=1$ in the linear theory.}

Similarly, the equal-time autocorrelation functions for {\em Case-II} are also known exactly in the linear theory. These are given by
\begin{subequations}
 \begin{align}
  \langle |h_1({\bf k})|^2 \rangle= \frac{2\tilde{D}\lambda^2_m}{\nu^2_\psi k^4}.\label{h1-corr}\\
 \langle |h_2({\bf k},t)|^2 \rangle= \frac{D}{\nu_\rho k^2}\label{h2-corr}
 \end{align}
\end{subequations}
in the Fourier space.
We thus find that $\chi_1=(4-d)/2$ and $\chi_2=(2-d)/2$ in linear theory; in this case $\mu_1=1=\mu_2$.  $z$ remains at 2  and $\chi_2$ has the same value in the linear theory in both the cases. 

\subsection{Nonlinear effects}\label{nonlin}
The nonlinear terms in (\ref{eqh1d}) and (\ref{eqh2d}) may change the scaling exponents found in the linear theory. The exponents can no longer be found exactly in the presence of the nonlinear terms. Furthermore, na\"ive perturbation theory produces diverging corrections to the model parameters. To handle these divergences systematically, we here use  perturbative dynamical RG approach to find the effects of anharmonic terms in scaling exponents for long wavelength limit.

The RG method is well documented in the literature \cite{chaikin, stanley, forster}; we briefly outline the steps of the RG analysis. The momentum shell dynamic RG consists of few steps, at first the fields are expressed as the sum of low and high wavevector modes: $h_1({\bf q})=h_1^<({\bf q})+h_1^>({\bf q})$ and $h_2({\bf q},\omega)=h_2^<({\bf q},\omega)+h_2^>({\bf q},\omega)$. Here $<$, $>$ present low and high wavevector modes respectively. Let the upper limit value of momentum is $\Lambda$ related to $1/a$, the microscopic cutoff in the system. We integrate out the high wavevector Fourier modes of fields restricted $\Lambda e^{-\delta l}<|{\bf q}|<\Lambda$, these are represented diagrammatically by the Feynman diagrams which are given in Appendix~\ref{perturbation theory}. This integration reduces the upper cut off of wavevector to $\Lambda e^{-\delta l}$ from $\Lambda$, and then we restore the upper limit of wavevector to $\Lambda$ by rescaling of  momentum by ${\bf q'}={\bf q}e^{\delta l}$, along with frequency by $\omega'=\omega e^{z\delta l}$, where $z$ is the dynamic exponent. The long wavelength parts of the fields are also rescaled; the necessary details are available in Appendix~\ref{perturbation theory}. This procedure ultimately yields the RG flow equations, which in turn gives the RG fixed points. 

\subsection{Universality class of Case-I}\label{caseI}

In this case, the surface is anisotropic due to the existence of the travelling waves with nonzero $\lambda_{1\psi}$, $\lambda_{1\rho}$. We write the model equations keeping the most relevant terms in the long wavelength limit below.
\begin{eqnarray}
 && -\nu_\psi\nabla^2h_1 - \lambda_{1\psi} \partial_\parallel h_1 - \lambda_m \delta m + \frac{\lambda_{\psi}}{2} ( \boldsymbol\nabla h_1)^2 \nonumber \\&&- \lambda_{2\psi}[\delta m (\partial_\parallel h_1)]  = 0,\label{eqh1d-nothalf}
\end{eqnarray}
and 
\begin{eqnarray}
 \frac{\partial h_2}{\partial t} &=& \nu_{\rho} \nabla^2 h_2 + \lambda_{1\rho} \partial_\parallel h_2 - \frac{\lambda_{\rho}}{2} ( \boldsymbol\nabla h_2)^2 + \lambda_{2\rho}[\delta m (\partial_\parallel h_2)]   \nonumber \\&-&  \lambda_{\rho\psi}( \boldsymbol\nabla h_1)\cdot ( \boldsymbol\nabla h_2) + f.\label{eqh2d-nothalf}
\end{eqnarray}
The perturbative corrections of these parameters  which  involve disorder (proportional to $\tilde D$) are finite; infrared divergent  corrections are due to the pure KPZ nonlinear term ($\frac{\lambda_{\rho}}{2} (\nabla h_2)^2$) only, see Appendix \ref{RG caseI}. Now to determine the scaling of the dimensionless effective coupling constants near the Gaussian fixed point, that depend upon  the disorder variance $\tilde D$. These  are  $a_0=\frac{\lambda_\psi^2 \tilde D \lambda_m^2}{\lambda_{1\psi}^4}$, $a_1=\frac{\tilde D \lambda_{2\psi}^2}{\lambda_{1\psi}^2}$, $a_2=\frac{\lambda_{\rho\psi}^2 \tilde D \lambda_m^2}{\lambda_{1\rho}^2\lambda_{1\psi}^2}$ , $a_3=\frac{\lambda_{2\rho}^2\tilde D}{\lambda_{1\rho}^2}$, $a_4=\frac{\lambda_\psi\lambda_{2\psi}\tilde{D}\lambda_m}{\lambda^3_{1\psi}}$. Under the rescaling of the momentum, frequency and the fields, the flow of these near the Gaussian fixed point~\cite{astik-prr}
\begin{eqnarray}
 &&\frac{da_0}{dl}= -da_0,\, \frac{da_1}{dl}= -da_1,\nonumber\\
 &&\frac{da_2}{dl}= -da_2,\, \frac{da_3}{dl}= -da_3,\, \frac{da_4}{dl}= -da_4.
\end{eqnarray}
Thus all these coupling constants are irrelevant near the Gaussian fixed point at all dimensions. The Galilean invariance  of the dynamics of $h_2({\bf x}, t)$ is restored in the long wavelength limit due to the irrelevance of the quenched disorder. The  flow of the dimensionless coupling constant $\tilde g=\frac{\lambda_\rho^2 D}{\nu_\rho^3}$, which appears in the pure KPZ problem as well, near the Gaussian fixed point is
\begin{equation}
 \frac{d\tilde g}{dl}= \tilde g [2-d].
\end{equation}
This is unstable for $d<2$, but stable for $d>2$. At any rate, even for $d>2$, $\tilde g$ remains more relevant near the Gaussian fixed point than the other dimensionless coupling constants as defined above, all of which have their origins in the disorder. We can thus conclude that disorder is irrelevant (in a RG sense) in the long wavelength limit at all dimensions for {\em Case-I}. This immediately implies that the long wavelength scaling is governed by the KPZ nonlinearity with the universal scaling belonging to the KPZ universality class. Thus, a roughening transition is expected at $d>2$, just as it is for the pure KPZ equation, to a rough phase identical to that for the pure KPZ equation, whose scaling properties are not accessible to the standard perturbative calculations. Furthermore and related to what we have just concluded, isotropy is restored in the long wavelength limit of the fluctuations of $h_2$. This implies $\mu_2=1$ in the renormalised equation, as in the pure KPZ equation.
This generalises one of the conclusions of Ref.~\cite{astik-prr} valid for $1d$. We therefore conclude that while the Gaussian fixed point  is stable with respect to perturbations by the quenched disorder, the pure KPZ nonlinearity  (with coupling $\tilde g$) remains relevant and determines the universal scaling properties. Thus the long wavelength properties of the model is given by the KPZ universality class. Therefore lower critical dimension is 2 and a roughening transition (i.e., a smooth-to-rough transition) exists as in the pure KPZ equation. { The upper critical dimension of Case-I, identical to that for the pure KPZ equation, is not known. It is speculated to be 4 by some authors~\cite{lassig97, jkb98, moser91, marinari02, wiese98}, although it remains debatable till the date.}

\subsection{Universality class of Case-II}\label{caseII}

We now discuss the universal scaling properties of Case-II.
In this case, the underdamped propagating modes vanish (since $\lambda_1\rho=0$) in the dynamics of $h_2({\bf x},t)$  as discussed before in Sec. \ref{Model}. Thus the surface is isotropic. We write down below the model equations keeping most relevant nonlinear terms:

\begin{eqnarray}
 -\nu_\psi \nabla^2 h_1 - \lambda_m \delta m + \frac{\lambda_\psi}{2} ( \boldsymbol\nabla h_1)^2 =0.\label{eqh1d-half}
\end{eqnarray}

\begin{eqnarray}
 \frac{\partial h_2}{\partial t} = \nu_{\rho} \nabla^2 h_2 -\lambda_{\rho\psi} [ \boldsymbol\nabla h_1 \cdot  \boldsymbol\nabla h_2] + f.\label{eqh2d-half}
\end{eqnarray}

The na\"ive perturbative fluctuation corrections to the  model parameters diverge in the long wavelength limit, which can be systematically handled within the RG framework. Here we perform  one loop   RG analysis; fluctuation corrections of parameters are represented by the one loop Feynman diagrams  whose details are available in Appendix \ref{RG caseII}. 
The flow equations of the model parameters are 
\begin{subequations}
 \begin{align}
   & \frac{d\nu_\psi}{dl}=\nu_\psi \left[d-2+(\frac{2}{d}-1)g_1 \right].\label{nu-psi flow}\\
 & \frac{d\lambda_m}{dl}=\lambda_m \left[-\chi_1-\frac{3d}{2}+\frac{g_1}{2} \right].\label{lambda-m flow}\\
 & \frac{d\lambda_\psi}{dl}=\lambda_\psi \left[-d-2+\chi_1-\frac{2g_1}{d} \right].\label{lambda-psi}\\
 & \frac{d\nu_\rho}{dl}=\nu_\rho \left[z-2+(\frac{2}{d}-1)g_2 \right].\label{nu-rho flow}\\
 & \frac{dD}{dl}=D \left[-d+z-2\chi_2+2g_2 \right].\label{D flow}\\
 & \frac{d \lambda_{\rho\psi}}{dl}=\lambda_{\rho\psi} \left[z-2+\chi_1-\frac{2g_2}{d} \right].\label{lambda-rhopsi flow}
 \end{align}
\end{subequations}
Here, the dimensionless coupling constants are $g_1=\frac{\lambda^2_\psi \tilde{D} \lambda^2_m}{\nu^4_\psi}\frac{\Omega_d}{(2\pi)^4}$, $g_2=\frac{\lambda^2_{\rho\psi} \tilde{D} \lambda^2_m}{\nu^2_\psi \nu^2_\rho}\frac{\Omega_d}{(2\pi)^4}$ involving $\tilde D$, where $\Omega_d$ is angular part of the $d$-dimensional loop integrals.
The flow equations of $g_1$ and $g_2$ are
\begin{eqnarray}
 &&\frac{dg_1}{dl}=g_1[4-d+(5-12/d)g_1],\label{g1 flow}\\
 &&\frac{dg_2}{dl}=g_2[4-d+(2-8/d)g_2+(3-4/d)g_1].\label{g3 flow}
\end{eqnarray}
{ Flow equations (\ref{g1 flow}) and (\ref{g3 flow}) clearly reveal that $d=4$ is the critical dimension of both $g_1$ and $g_2$.}
Let us now calculate the fixed point (FP) values of $g_1$ and $g_2$ at different dimensions by setting $dg_1/dl=0=dg_2/dl$. { The Gaussian FP is $g_1^*=0=g_2^*$, furthermore one can find easily  nontrivial FP as $g_1^*=\frac{4-d}{12/d-5}$   from (\ref{g1 flow}) and using this in (\ref{g3 flow}) $g_2^*=\frac{4-d}{12/d-5}$ same as $g_1^*$, shown in Fig. \ref{phase}. Notice that the nontrivial fixed point diverges at $d=12/5$ and physically unacceptable between the dimensions $12/5$ and $4$.
\begin{figure}
 \includegraphics[scale=0.8]{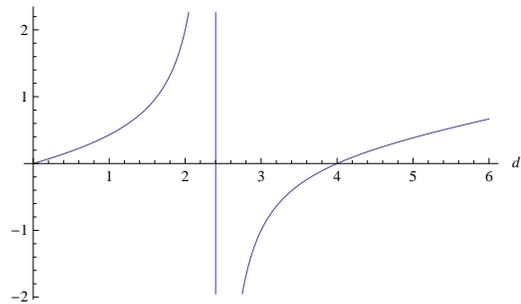}
\caption{Nontrivial fixed point value of $g_1$ as a function of $d$ (surface dimension) in one loop calculation. This is also same for $g_2$; see text.}\label{phase}
\end{figure}

We find in detail below for different dimensions:

 i) $d\leqslant12/5$: In this regime the Gaussian FP $g_1^*=0=g_2^*$ is  unstable. Thus $g_1^*$ and $g_2^*$ should be {\em non-zero} in the statistical steady state, implying relevance (in a RG sense) of the quenched disorder. The nontrivial FP is linearly stable.   The scaling exponents at the stable fixed point are given below.
 
 (a) $d=1$: The exponents are $\chi_1=9/7$, $\chi_2=5/7$, $z=11/7$. Thus  $\chi_2+z=16/7>2$, in contrast to the pure $1d$ KPZ equation. The results are consistent with \cite{astik-prr}.
 
 (b) $d=2$: The exponents are $\chi_1=2$, $\chi_2=2$, $z=2$. Also here $\chi_2+z>2$, unlike the pure KPZ equation. }

 (ii) Dimension $12/5\leqslant d<4 $: For this intermediate range of dimensions, physically acceptable fixed point is only $g_1^*=0$, $g_2^*=0$, which however is an unstable fixed point. The perturbation theory apparently breaks down for $12/5\leqslant d<4 $, which we believe is an artifact of the one-loop expansion employed here. This is similar to what we find for $3/2<d<2$ in the one-loop RG calculations for the pure KPZ equation; see Eq.~(\ref{kpz-flow})~ \cite{Frey-two-loop}. A schematic RG flow diagram is shown in Fig.~\ref{flow1} below.
 \begin{figure}[htb]
 \includegraphics[width=6cm]{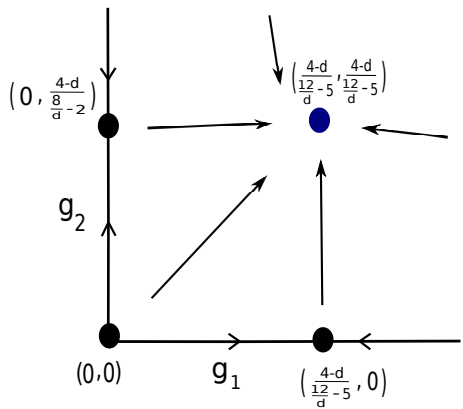}
  \caption{Schematic RG flow diagram for $d<4$. Stable and unstable fixed points are shown; flow lines indicate the direction of the RG flows towards stable FP (see text).}\label{flow1}
 \end{figure}

 (iii) Dimension $d=4$: It is the lower critical dimension of the presented model here. The nonlinear couplings are marginally relevant, are shown in (\ref{g1 flow}) and (\ref{g3 flow}). This implies that at $d=4$, the coupling constants flow to infinity in a finite RG ``time'' or for a finite system size, that is controlled by the precise values of the unrenormalised model parameters (and hence non-universal). This is reminiscent of the flow of the coupling constant at $2d$ in the pure KPZ equation. 

 (iv) Dimension $d> 4$: Let $d=4+\epsilon$, where $\epsilon>0$ is assumed to be small. The Gaussian fixed points are now stable, corresponding to $z=2$, $\chi_1=2-d/2$, and $\chi_2=1-d/2$, corresponding to asymptotically smooth phases.  In addition, similar to the pure KPZ equation, we expect unstable fixed points with both $g_1^*,\,g_2^*\sim {\cal O}(\epsilon)$, that would indicate a disorder induced roughening transition in the model. In fact, we do find $g_1^*=\epsilon/2$. This however leaves $g_2$ marginal. However, given the fact that $g_2^*=0$ is a stable fixed point (along with $g_1^*=0)$, and $g_1^*$ has an unstable fixed point that is ${\cal O}(\epsilon)$, we speculate that $g_2$ too shows a roughening transition such that for large enough bare or unrenormalised value of $g_2$, it flows to infinity under renormalisation. Our one-loop RG appears to be inadequate to capture this behaviour; higher order perturbation theory or numerical solutions of the dynamical equation could be useful in this regard. This further indicates that $d=4$ is the lower critical dimension for both $g_1$ and $g_2$.

  { Some studies \cite{lassig97, jkb98, moser91, marinari02, wiese98} suggest $d=4$ as the upper critical dimension on or above which the scaling exponents from the linear theory holds, for the pure KPZ equation. But here $d=4$ is found to be the  lower critical dimension, which is higher than the lower critical dimension of the pure KPZ equation. Since the upper critical dimension of a model should be higher than the lower critical dimension,  we expect the upper critical dimension for Case II should be greater than $d=4$. }
  
\begin{figure}[htb]
\includegraphics[width=6cm]{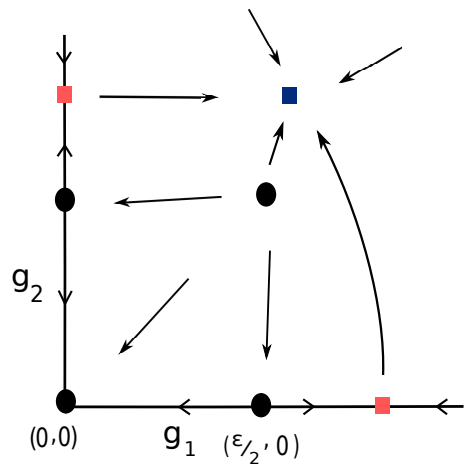}
\caption{Speculated global RG flows in the $g_1-g_2$ plane in an Occam's razor style argument. Perturbative accessible (small filled black circles) and inaccessible (small coloured squares) are shown. RG flow lines with directions are marked. Notice not all the unstable and in-principle perturbatively accessible fixed points are captured by the one-loop theory; see text.}\label{flow2}
 
\end{figure}

  { In the pure KPZ equation, the stable fixed point that governs the rough phase is inaccessible in a perturbation theory, although such a fixed point should exist on physical ground. Similarly, here too a globally stable fixed point that characterises the rough phase must exist. Using an ``Occam's razor'' style argument, we speculate the RG flow diagram Fig. \ref{flow2} that has the physically acceptable and simplest flow topology. Notice that at this putative strong coupling fixed point, both $g_1$ and $g_2$ are expected to be non-zero, for if $g_2=0$ at this fixed point, the rough phase should become identical to that of the pure KPZ equation. Such a possibility is however ruled out, since a strong coupling perturbatively inaccessible rough phase appears already at $2d$ for the pure KPZ equation, whereas for Case-II here, it can appear only at $4d$ or above. Thus the rough phase here is expected to be different from its counterpart for the pure KPZ equation.  }

Finally, we give a schematic, pictorial comparison between the phases and phase transitions in Case-I and Case-II as a function of dimension $d$ in Fig.~\ref{dim1} below.
\begin{figure}[htb]
\includegraphics[width=6cm]{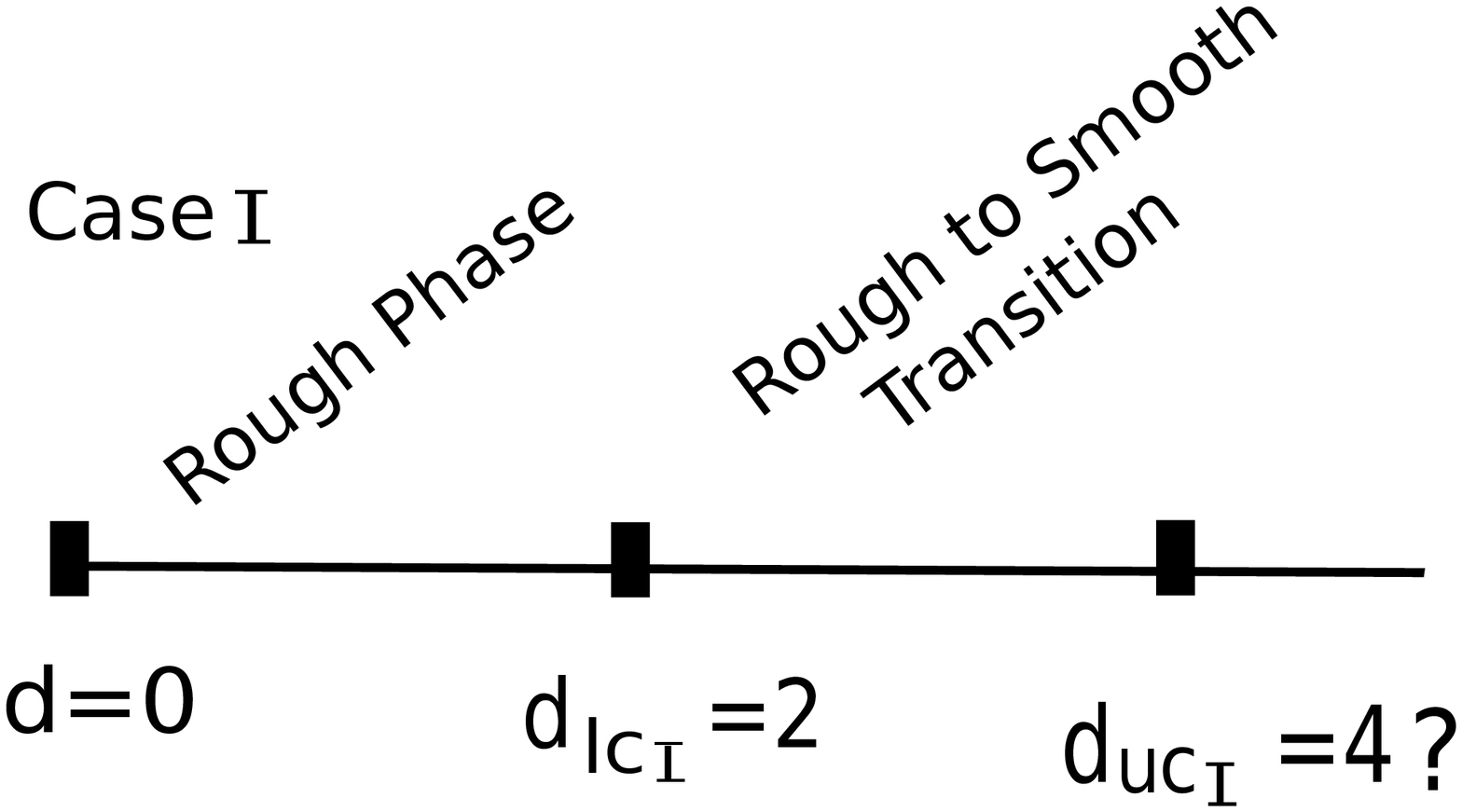}\\
\includegraphics[width=6cm]{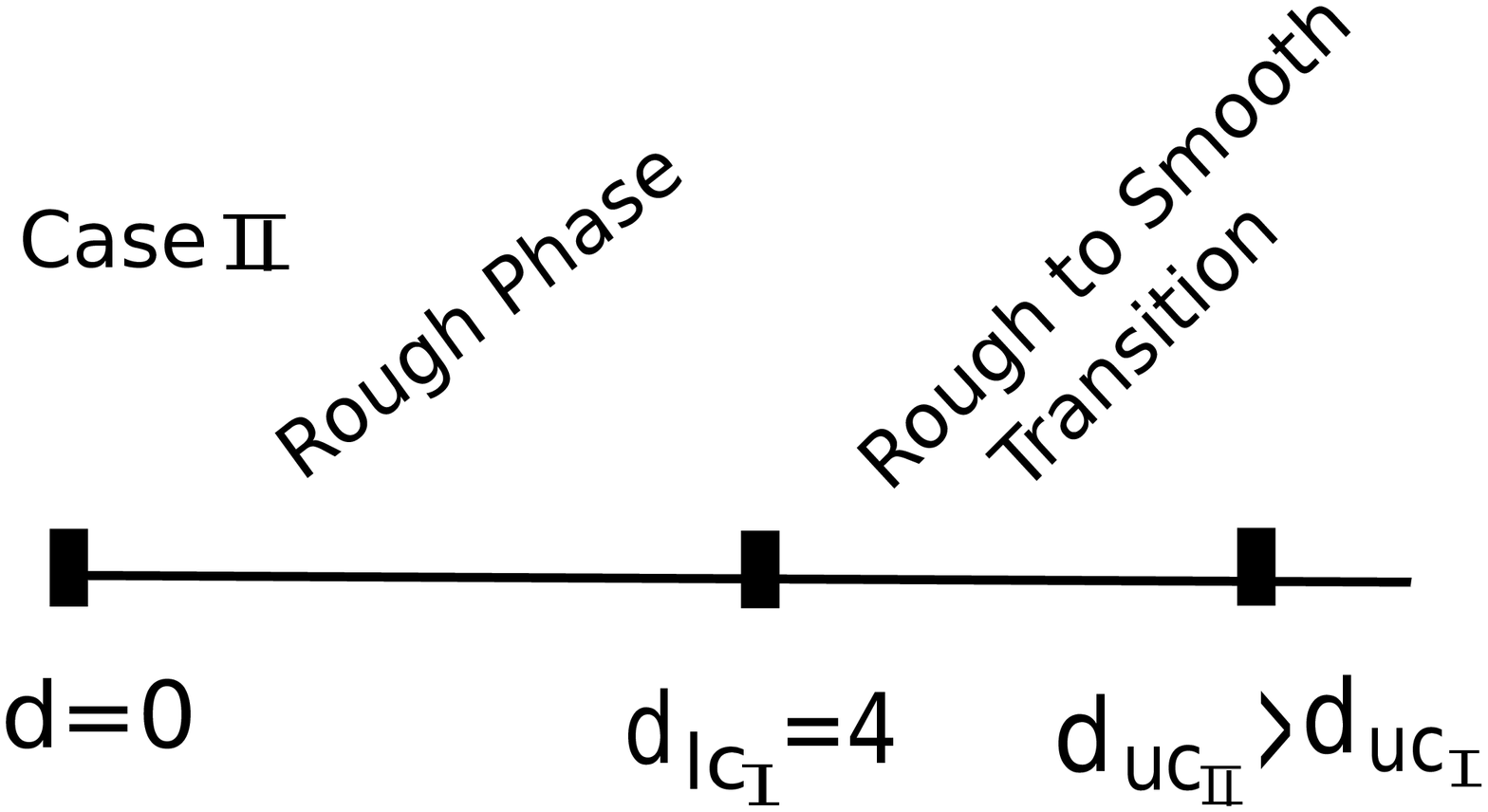}
\caption{Phases and phase transitions in the model as a function of $d$. (top) Case-I and (bottom) Case-II. Lower ($d_{lc_I}=2$) and ($d_{lc_{II}}=4$) dimensions for Case-I and Case-II, respectively,  and (yet unknown) upper critical ($d_{uc_I}$) and $d_{uc_{II}}$ dimensions, respectively, for  Case-I and Case-II  are marked. Further, $d_{uc_{II}}$ is expected to be higher than $d_{uc_I}$; see text.}\label{dim1}
\end{figure}

{ We expect the new universality class that emerges in the absence of the propagating waves holds not only when these waves strictly vanish, but also when these have ``small'' amplitudes, i.e., a nonuniversal ``small'' window of speeds of propagation around zero, for reasons identical to those enunciated in Ref.~\cite{astik-prr}.  Essentially for sufficiently high wavevectors, the propagating waves are subdominant to the damping terms. If under mode elimination from the high wavevector limits, the theory gets renormalised {\em before}, the crossover scale $\xi_c$ between the propagating modes and damping terms, the ``renormalised'' theory at the crossover scale would like what we have obtained for Case-II. Since at length scales larger than $\xi_c$, the propagating modes are important, and no further renormalisation takes place. As a result, the scaling behaviour corresponding to our Case-II above will be displayed by the system, with a non-KPZ scaling behaviour. If, on the other hand, the theory does not get ``renormalised'' till the crossover scale $\xi_c$ is reached, further renormalisation of the theory at scales larger than $\xi_c$ takes place, which comes entirely from the KPZ nonlinearity. This naturally leads to a scaling behaviour identical to the pure KPZ   equation; see Ref.~\cite{astik-prr} for more detailed discussions. }

\section{Summary and outlook}\label{summary}

To summarise, we have studied here the universal scaling properties of $d$-dimensional KPZ equation with quenched columnar disorder. For this, we have generalised the $1d$ continuum equation constructed and studied in Ref.~\cite{astik-prr}. We have obtained a number of interesting results. For instance, we show that the columnar disorder in general leads to the loss of the Galilean invariance of the pure KPZ equation and generation of underdamped propagating waves in the system, which in turn makes the system anisotropic. Interestingly, these waves render the quenched disorder irrelevant (in a RG sense); as a result, the universal scaling exponents belong to the $d$-dimensional KPZ universality class. Since the pure KPZ equation is isotropic, this implies that the long wavelength scaling properties of the model are actually isotropic. { We argue that the rough phase of the model that is inaccessible in a perturbation theory, is statistically identical to that for the pure KPZ equation in the long wavelength limit.} Thus isotropy becomes an {\em effective} symmetry in the long wavelength limit. For certain choices of the model parameters, the propagating waves vanish. In that limit the model is already isotropic. Furthermore, the quenched disorder in the absence of the waves is now {\em relevant} (in a RG sense). We show that the model now has $d=4$ as the {\em lower critical dimension}. We have calculated the roughness and the dynamic scaling exponents within a one-loop approximation, which belong to a universality class hitherto unknown. We argue that above $d=4$ the model undergoes a roughening transition from a smooth to rough phase that is analogous to the well-known roughening transition in the KPZ equation above $d=2$. { This rough phase, although not accessible in a perturbation theory, is statistically {\em different} from its counterpart in the pure KPZ equation.} We have argued that the upper critical dimension in Case II should be higher than 4.

We have considered only Gaussian-distributed short-ranged quenched disorder. Our calculational framework can be extended to Gaussian-distributed long-range disorder in a straight forward manner. While precise values of the scaling exponents should depend upon the scaling of the variance of the long-range disorder, by using the logic outlined above we can generally argue that the universal scaling properties will crucially depend upon whether or not there are underdamped propagating waves. We expect our results here will provide impetus to future theoretical work along this direction.

\section{acknowledgements}

The author would like to acknowledge Prof. Abhik Basu (SINP, Kolkata) for helpful discussions, valuable suggestions and critical comments to prepare the  manuscript.

\appendix
\section{Perturbation Theory: Generating Functional and Scaling}\label{perturbation theory}

The Generating Functional \cite{zinn, tauber, dedominicis, bausch} is defined as
\begin{equation}
 {\mathcal{Z}}=\int {\cal D}\hat{h}_1 {\cal D}h_1 {\cal D}\hat{h}_2 {\cal D}h_2 {\cal D} \delta m \exp\{-{\cal S}[\hat{h}_1, h_1, \hat{h}_2, h_2, \delta m]\}.
\end{equation}
$\hat{h}_1,\, \hat{h}_2$ are the conjugate fields to $h_1,\, h_2$ respectively. Here, ${\cal S}[\hat{h}_1, h_1, \hat{h}_2, h_2, \delta m]$ is the action functional. Two types of terms are present in the action ${\cal S}$: linear terms and nonlinear terms. The perturbation theory is set up by expanding the nonlinear terms present in the action.

 The perturbative step of evaluating fluctuation corrections is followed by rescaling  momentum (equivalently, space) and frequency (equivalently, time)  as ${\bf q}\rightarrow b{\bf  q}$ and $\omega\rightarrow b^z\omega$ respectively. The long wavelength parts of the fields are rescaled as follows:
\begin{eqnarray}
&&\hat h_1({\bf q})=b^{-\chi_1-d}\hat h_1(b{\bf q}),\;h_1({\bf q})=b^{d+\chi_1}h_1(b{\bf q}),\nonumber \\&&\delta m({\bf q})=b^{d/2}\delta m(b{\bf q}),\nonumber\\ 
&&\hat h_2({\bf q},\omega)=b^{z-\chi_2}\hat h_2(b{\bf q},b^z\omega),\nonumber \\&& h_2({\bf q},\omega)=b^{d+z+\chi_2}h_2(b{\bf q},b^z\omega). \label{rescaling11}
\end{eqnarray}

\section{Renormalisation Group analysis for Case-I}\label{RG caseI}

The action functional corresponding to equations (\ref{eqh1d-nothalf}) and (\ref{eqh2d-nothalf}) is
\begin{eqnarray}
 &&{\cal S}[\hat{h}_1, h_1, \hat{h}_2, h_2, \delta m] = \int_{\bf x} \hat{h}_1[-\nu_\psi\nabla^2h_1 - \lambda_{1\psi} \partial_\parallel h_1 \nonumber\\&&- \lambda_m \delta m + \frac{\lambda_{\psi}}{2} (\boldsymbol\nabla h_1)^2 - \lambda_{2\psi}[\delta m (\partial_\parallel h_1)]+\frac{\delta m^2}{4\tilde{D}} \nonumber\\ 
 &&+\int_{{\bf x}, t} \hat{h}_1[-D\hat{h}_2+\partial_t h_2- \nu_{\rho} \nabla^2 h_2 -\lambda_{1\rho} \partial_\parallel h_2 + \frac{\lambda_{\rho}}{2} (\boldsymbol\nabla h_2)^2 \nonumber\\&& - \lambda_{2\rho}[\delta m (\partial_\parallel h_2)]   + \lambda_{\rho\psi}(\boldsymbol\nabla h_1)\cdot (\boldsymbol\nabla h_2)] .\label{action-I}
\end{eqnarray}

\begin{figure}
\includegraphics[scale=0.4]{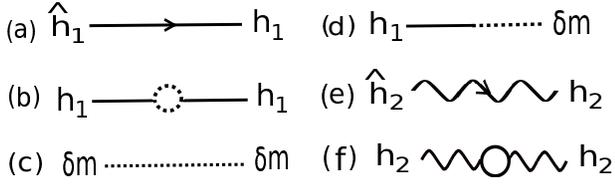}
\caption{Diagramatic representations of two point functions. Arrows in propagators ((a), (e)) constitute casuality information.}\label{correlators}
\end{figure}

\begin{figure}
 \includegraphics[scale=0.4]{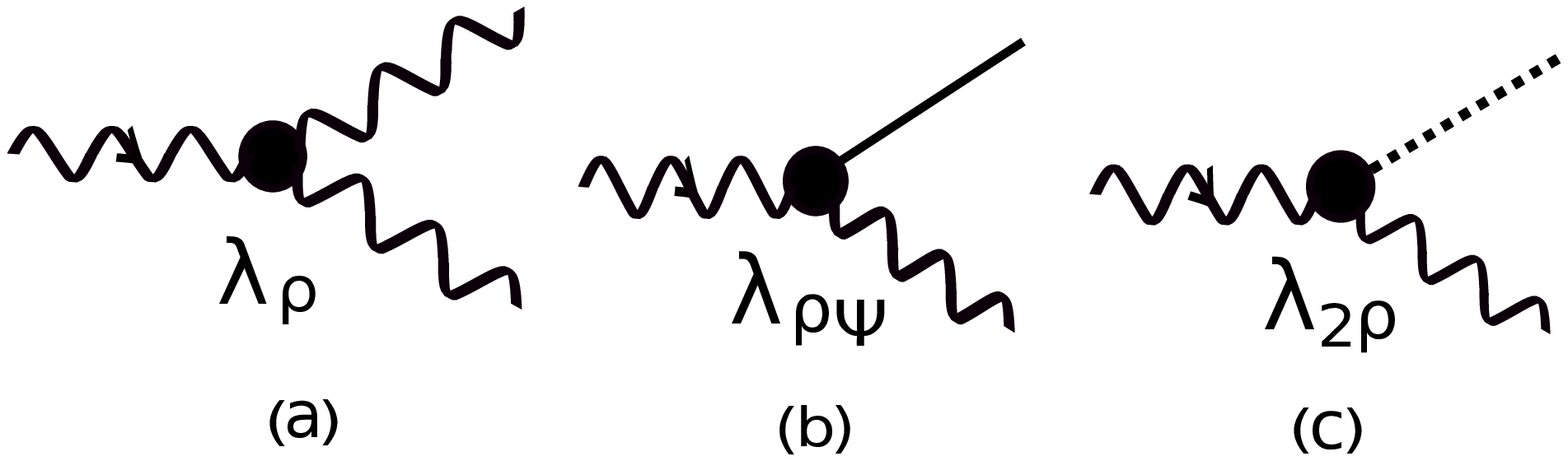}
 \includegraphics[scale=0.4]{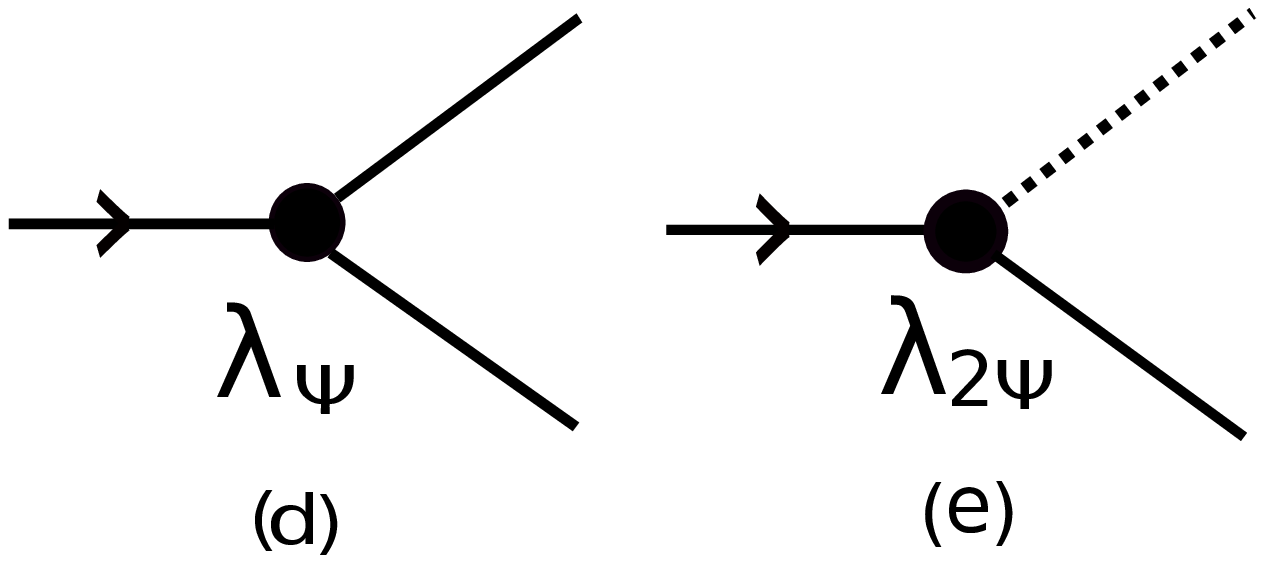}
\caption{Diagramatic representations of the  anharmonic vertices terms in action~\ref{action-I}.}\label{vertex}
\end{figure}

The two point functions found from linear terms present in (\ref{action-I}) and nonlinear vertices in (\ref{action-I}) are presented diagramatically in Fig. \ref{correlators} and Fig. \ref{vertex} respectively.

The bare two point functions in the harmonic theory neglecting nonlinear terms from (\ref{action-I}) are
\begin{subequations}
 \begin{align}
  &\langle \hat{h}_1(-{\bf k}) h_1({\bf k}) \rangle = \frac{1}{\nu_\psi k^2+ik_\parallel\lambda_{1\psi}}.\\
 &\langle | h_1({\bf k})|^2 \rangle = \frac{2\tilde D \lambda_m^2}{k_\parallel^2\lambda_{1\psi}^2+\nu_\psi^2 k^4}.\\
 &\langle h_1(-{\bf k}) \delta m ({\bf k}) \rangle =\frac{2\tilde D \lambda_m}{\nu_\psi k^2-ik_\parallel\lambda_{1\psi}}.\\
 &\langle \hat{h}_2(-{\bf k},-\omega) h_2({\bf k},\omega) \rangle = \frac{1}{-i\omega+ik_\parallel\lambda_{1\rho}+ \nu_\rho k^2}.\\
 &\langle |h_2({\bf k},\omega)|^2 \rangle = \frac{2D}{(\omega-k_\parallel\lambda_{1\rho})^2+ \nu_\rho^2 k^4}.
 \end{align}
\end{subequations}

\begin{figure}
\includegraphics[scale=0.4]{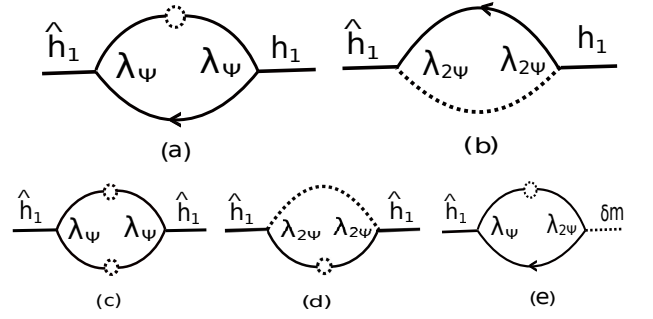}
\caption{One loop Feynman diagrams for the corrections of the propagator and correlator of $h_1$: (a),(b) for propagator and (c),(d),(e) for correlator corrections. }\label{h1-corrections}
\end{figure}

\begin{figure}
\includegraphics[scale=0.4]{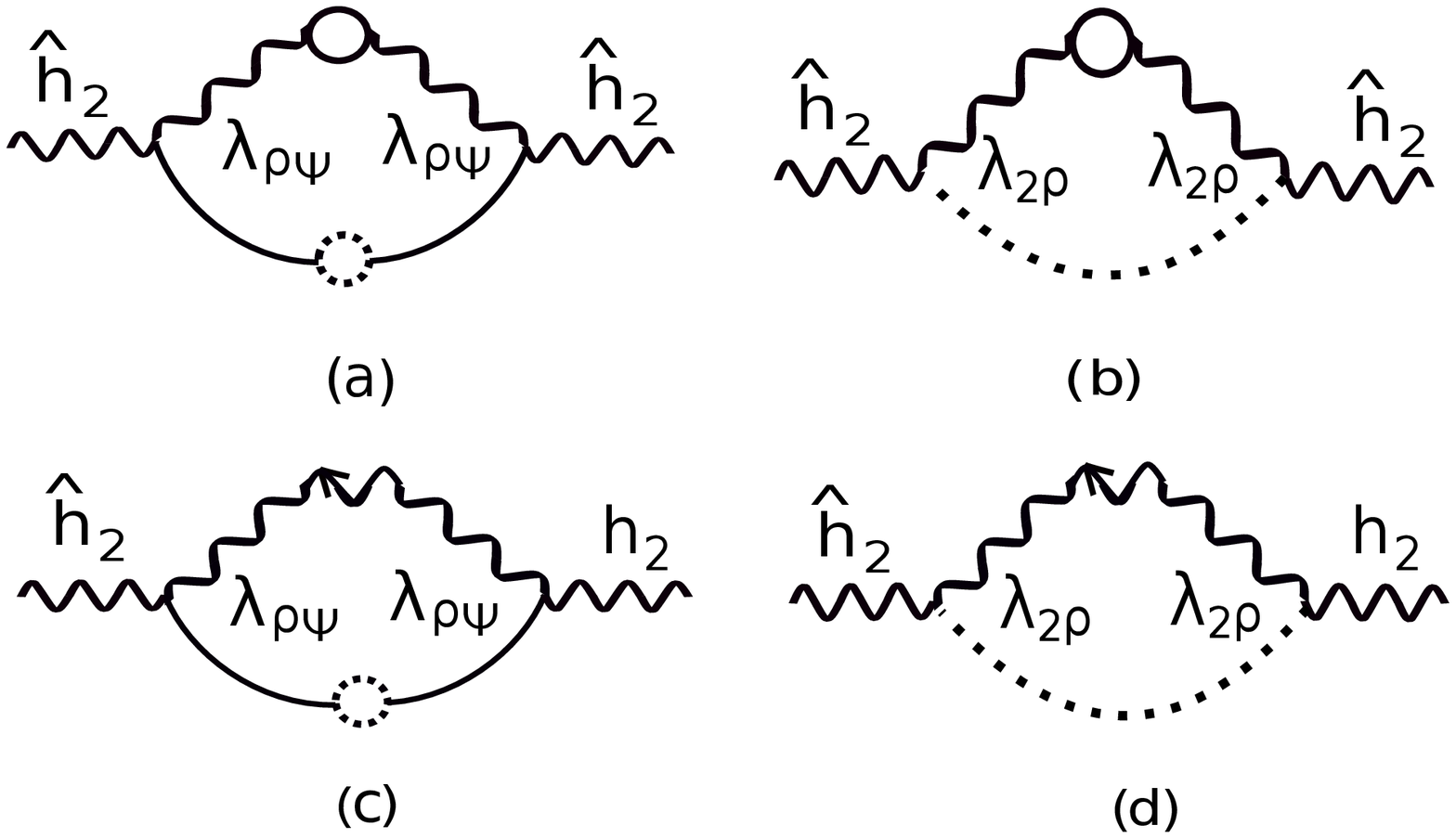}
\caption{One loop Feynman diagrams for the corrections of the correlator and propagator of $h_2$: (a),(b) for correlator and (c),(d) for propagator corrections.}\label{h2-corrections}
\end{figure}

The fluctuation corrections of the propagators and correlators of $h_1,\, h_2$ are represented by one-loop Feynman diagrams in Fig. \ref{h1-corrections} and Fig. \ref{h2-corrections} respectively.

The one loop contributions to the corrections of parameters in (\ref{action-I}) are finite. Those are of the forms
\begin{subequations}
 \begin{align}
  {\rm Fig.} (\ref{h1-corrections}a) &=  \frac{\lambda^2_\psi \tilde{D} \lambda^2_m}{i \lambda^3_{1\psi} } \int \frac{d^d{\bf q}}{(2\pi)^d} \frac{1}{q^3_\parallel}[k^2q^2+({\bf k}\cdot{\bf q})^2\nonumber\\   &-q^2{\bf k}\cdot{\bf q}(2+3k_\parallel/q_\parallel)].\\
 {\rm Fig.} (\ref{h1-corrections}b) &= \frac{2\lambda^2_{2\psi}\tilde{D}}{\lambda_{1\psi}} \int \frac{d^d{\bf q}}{(2\pi)^d}.\\
  {\rm Fig.} (\ref{h1-corrections}c) &= \frac{\lambda^2_{\psi}(\tilde{D}\lambda^2_m)^2}{\lambda^4_{1\psi}} \int \frac{d^d{\bf q}}{(2\pi)^d}.\\
  {\rm Fig.} (\ref{h1-corrections}d) &= \frac{2\lambda^2_{2\psi}\tilde{D}^2\lambda^2_m}{\lambda^2_{1\psi}} \int \frac{d^d{\bf q}}{(2\pi)^d}.\\
  {\rm Fig.} (\ref{h1-corrections}e) &= \frac{2\lambda_\psi\lambda_{2\psi}\tilde{D}\lambda^2_m}{\lambda^3_{1\psi}} \int \frac{d^d{\bf q}}{(2\pi)^d} \frac{q^2}{q^2_\parallel}.
 \end{align}
\end{subequations}

\begin{subequations}
 \begin{align}
{\rm Fig.} (\ref{h2-corrections}a) &= \frac{2D\tilde{D}\lambda_{\rho\psi}^2\lambda_m^2}{\lambda^2_{1\rho}\lambda^2_{1\psi}} \int \frac{d^d{\bf q}}{(2\pi)^d} \frac{q^2}{q^2_\parallel}.\\
 {\rm Fig.} (\ref{h2-corrections}b) &= -\frac{2D\lambda^2_{2\rho}}{\lambda^2_{1\rho}}\int \frac{d^d{\bf q}}{(2\pi)^d} \frac{q^2_\parallel}{q^2}.\\
 {\rm Fig.} (\ref{h2-corrections}c) &= \frac{i\tilde{D}\lambda^2_{\rho\psi}\lambda^2_m}{\lambda_{1\rho}\lambda^2_{1\psi}}\int \frac{d^d{\bf q}}{(2\pi)^d} \frac{1}{qq^2-\parallel}[-2({\bf k}\cdot{\bf q})^2-k^2q^2]\nonumber\\ & ~~~~~~~~ +\frac{q({\bf k}\cdot{\bf q})}{q^2_\parallel}[2+k/q+k_\parallel/q_\parallel].
 \end{align}
 \end{subequations}

\section{Renormalisation Group analysis for Case-II}\label{RG caseII}

Action functional corresponding to equations (\ref{eqh1d-half}) and (\ref{eqh2d-half}) is
\begin{eqnarray}
 &&{\cal S}=\int_{\bf x} \hat{h}_1 [-\nu_\psi \nabla^2 h_1 - \lambda_m \delta m + \frac{\lambda_\psi}{2} (\boldsymbol\nabla h_1)^2] + \frac{\delta m^2}{4\tilde{D}} \nonumber\\
 && +\int_{{\bf x},t} \hat{h}_2[-D\hat{h_2} + (\partial_t h_2 - \nu_\rho \nabla^2 h_2 + \lambda_{\rho\psi}\boldsymbol\nabla h_1 \cdot \boldsymbol\nabla h_2)].\label{action-II}
\end{eqnarray}

Two point functions of the harmonic theory neglecting nonlinear terms fom (\ref{action-II}) are
\begin{subequations}
 \begin{align}
   &\langle \hat{h}_1(-{\bf k}) h_1({\bf k}) \rangle = \frac{1}{\nu_\psi k^2}.\\
 &\langle | h_1({\bf k})|^2 \rangle = \frac{2\tilde D \lambda_m^2}{\nu_\psi^2 k^4}.\\
 &\langle h_1(-{\bf k}) \delta m ({\bf k}) \rangle =\frac{2\tilde D \lambda_m}{\nu_\psi k^2}.\\
 &\langle \hat{h}_2(-{\bf k},-\omega) h_2({\bf k},\omega) \rangle = \frac{1}{-i\omega+ \nu_\rho k^2}.\\
 &\langle |h_2({\bf k},\omega)|^2 \rangle = \frac{2D}{\omega^2+ \nu_\rho^2 k^4}.
 \end{align}
\end{subequations}

The two point functions for this case  are also represented diagramatically in Fig.\ref{correlators}. For this case, the nonlinear vertices are presented by diagrams (\ref{vertex}a), (\ref{vertex}b) and (\ref{vertex}d) in Fig.\ref{vertex}.

\begin{figure}
\includegraphics[scale=0.4]{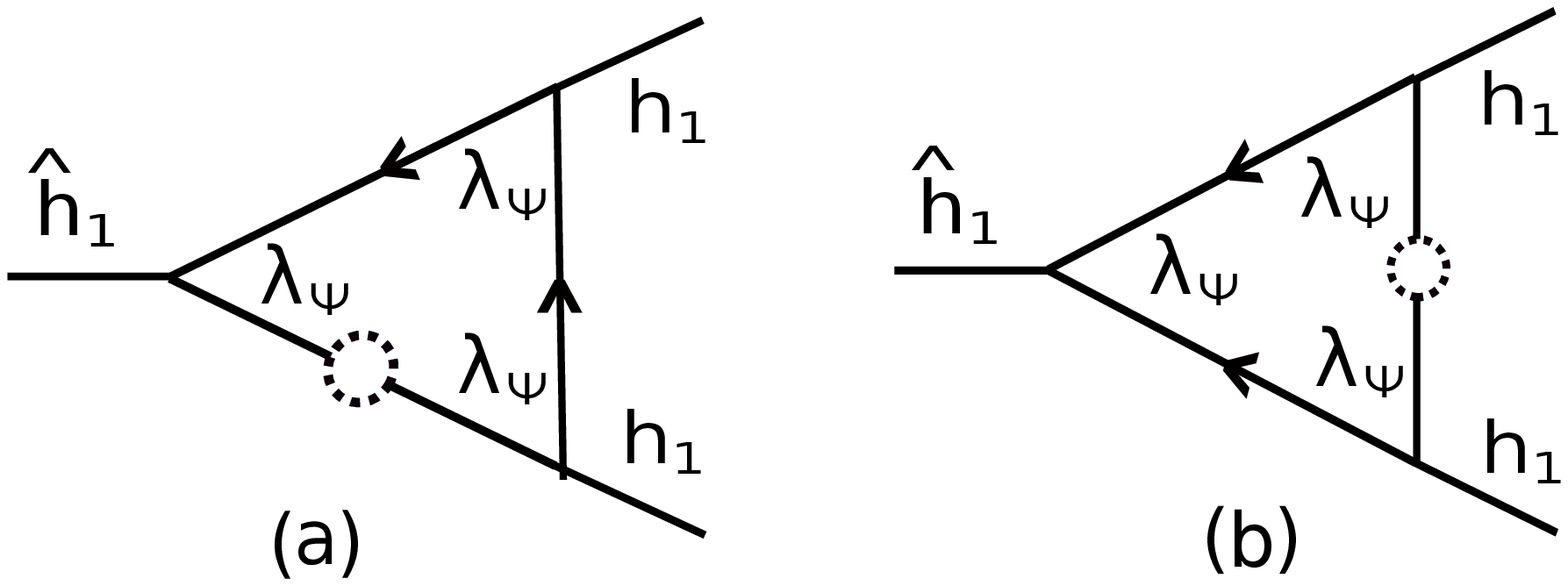}
\caption{One loop Feynman diagram for the vertex $\lambda_{\psi}$ correction for case-II.}\label{h1-vertex-correction}
\end{figure}

\begin{figure}
\includegraphics[scale=0.4]{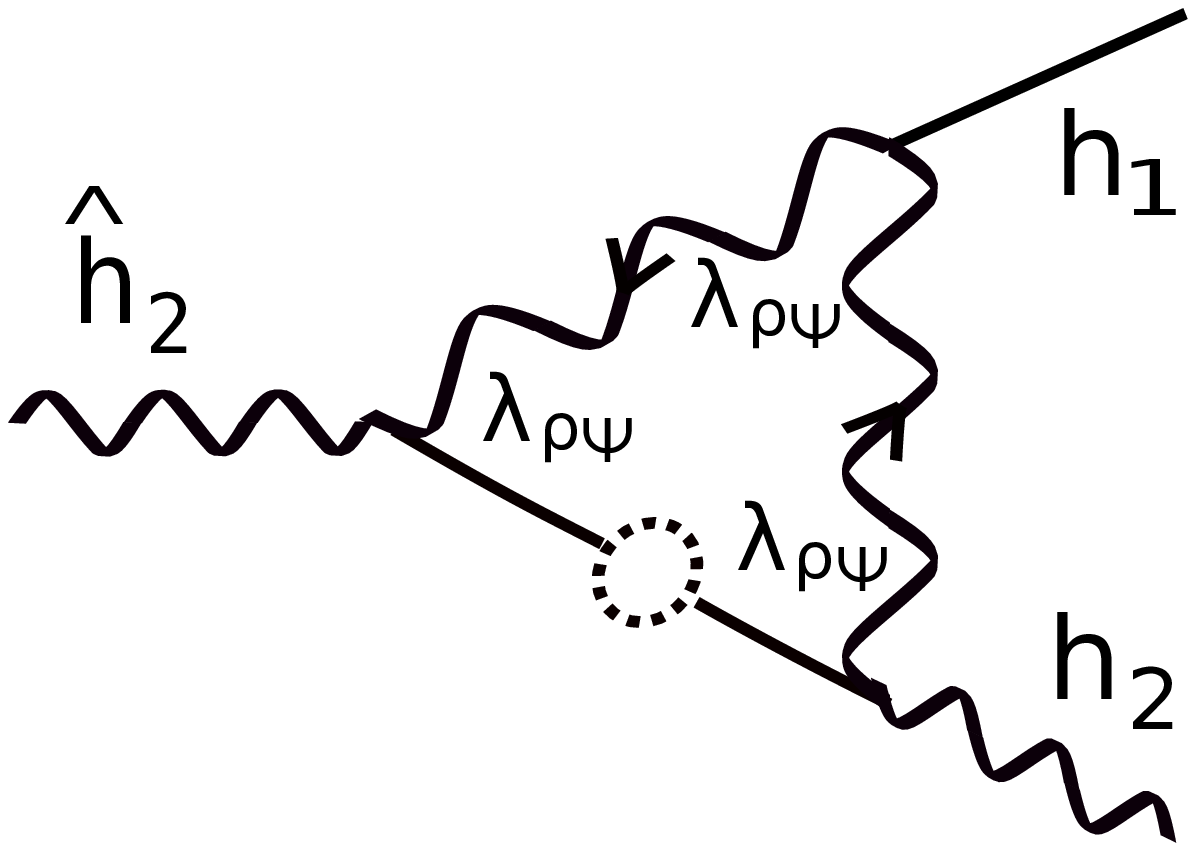}
\caption{One loop Feynman diagram for the vertex $\lambda_{\rho\psi}$ correction for case-II.}\label{h2-vertex-correction}
\end{figure}

In this case, the one loop diagrams in Fig.(\ref{h1-corrections}a), (\ref{h1-corrections}c) contribute to the corrections of the propagator and correlator of $h_1$ respectively. The diagrams in Fig.(\ref{h2-corrections}a), (\ref{h2-corrections}c) contribute to the corrections of correlator and propagator of $h_2$ respectively. Fig.\ref{h1-vertex-correction}, Fig.\ref{h2-vertex-correction} contribute to vertices corrections in (\ref{action-II}).

The one loop contributions to corrections of parameters in (\ref{action-II}) are
\begin{subequations}
 \begin{align}
  &{\rm Fig.} (\ref{h1-corrections}a) =  \frac{\lambda^2_\psi \tilde{D} \lambda^2_m}{\nu^3_\psi }[\frac{2}{d}-1] \int^{\varLambda}_{\varLambda/b} \frac{d^d{\bf q}}{(2\pi)^d} \frac{1}{q^4}.\\
  &{\rm Fig.} (\ref{h1-corrections}c)= \frac{\lambda^2_\psi \tilde{D}^2 \lambda^2_m}{\nu^4_\psi } \int^{\varLambda}_{\varLambda/b} \frac{d^d{\bf q}}{(2\pi)^d} \frac{1}{q^4}.
 \end{align}
\end{subequations}

\begin{subequations}
 \begin{align}
  &{\rm Fig.} (\ref{h2-corrections}a) =  \frac{2D\lambda^2_{\rho\psi} \tilde{D} \lambda^2_m}{\nu^2_\psi \nu^2_\rho} \int^{\varLambda}_{\varLambda/b} \frac{d^d{\bf q}}{(2\pi)^d} \frac{1}{q^4}.\\
  &{\rm Fig.} (\ref{h2-corrections}c)= \frac{\lambda^2_{\rho\psi} \tilde{D} \lambda^2_m}{\nu^2_\psi \nu_\rho }[1-2/d] \int^{\varLambda}_{\varLambda/b} \frac{d^d{\bf q}}{(2\pi)^d} \frac{1}{q^4}.
 \end{align}
\end{subequations}

\begin{subequations}
 \begin{align}
  &{\rm Fig.} (\ref{h1-vertex-correction}a)= \frac{2\lambda^3_\psi \tilde{D} \lambda^2_m}{\nu^4_\psi d} \int^{\varLambda}_{\varLambda/b} \frac{d^d{\bf q}}{(2\pi)^d} \frac{1}{q^4}. \\
  &{\rm Fig.} (\ref{h1-vertex-correction}b)= -\frac{\lambda^3_\psi \tilde{D} \lambda^2_m}{\nu^4_\psi d} \int^{\varLambda}_{\varLambda/b} \frac{d^d{\bf q}}{(2\pi)^d} \frac{1}{q^4}.
 \end{align}
\end{subequations}

\begin{equation}
 {\rm Fig.} (\ref{h2-vertex-correction})= \frac{2\lambda^3_{\rho\psi} \tilde{D} \lambda^2_m}{\nu^2_\psi \nu^2_\rho d} \int^{\varLambda}_{\varLambda/b} \frac{d^d{\bf q}}{(2\pi)^d} \frac{1}{q^4}.
\end{equation}

Finally one loop corrected parameters in (\ref{action-II}) are
\begin{subequations}
 \begin{align}
& \nu^<_\psi=\nu_\psi \left[1+(\frac{2}{d}-1)\frac{\lambda^2_\psi \tilde{D} \lambda^2_m}{\nu^4_\psi} \int^{\varLambda}_{\varLambda/b} \frac{d^d{\bf q}}{(2\pi)^d} \frac{1}{q^4} \right].\\
 & \lambda_m^<= \lambda_m \left[1+\frac{\lambda^2_\psi \tilde{D} \lambda^2_m}{2\nu^4_\psi} \int^{\varLambda}_{\varLambda/b} \frac{d^d{\bf q}}{(2\pi)^d} \frac{1}{q^4} \right].\\
& \lambda_\psi^<= \lambda_\psi \left[1-\frac{\lambda^2_\psi \tilde{D} \lambda^2_m}{\nu^4_\psi} \frac{2}{d}\int^{\varLambda}_{\varLambda/b} \frac{d^d{\bf q}}{(2\pi)^d} \frac{1}{q^4} \right].\\
& \nu^<_\rho=\nu_\rho \left[1+(\frac{2}{d}-1)\frac{\lambda^2_{\rho\psi} \tilde{D} \lambda^2_m}{\nu^2_\psi \nu^2_\rho} \int^{\varLambda}_{\varLambda/b} \frac{d^d{\bf q}}{(2\pi)^d} \frac{1}{q^4} \right].\\
& D^<=D \left[1+\frac{2\lambda^2_{\rho\psi} \tilde{D} \lambda^2_m}{\nu^2_\psi \nu^2_\rho} \int^{\varLambda}_{\varLambda/b} \frac{d^d{\bf q}}{(2\pi)^d} \frac{1}{q^4} \right].\\
& \lambda_{\rho\psi}^<= \lambda_{\rho\psi} \left[1-\frac{\lambda^2_{\rho\psi} \tilde{D} \lambda^2_m}{\nu^2_\psi \nu^2_\rho} \frac{2}{d}\int^{\varLambda}_{\varLambda/b} \frac{d^d{\bf q}}{(2\pi)^d} \frac{1}{q^4} \right].
 \end{align}
\end{subequations}


\bibliography{qkpzddim.bib}

\begin{thebibliography}{25}
\expandafter\ifx\csname natexlab\endcsname\relax\def\natexlab#1{#1}\fi
\expandafter\ifx\csname bibnamefont\endcsname\relax
  \def\bibnamefont#1{#1}\fi
\expandafter\ifx\csname bibfnamefont\endcsname\relax
  \def\bibfnamefont#1{#1}\fi
\expandafter\ifx\csname citenamefont\endcsname\relax
  \def\citenamefont#1{#1}\fi
\expandafter\ifx\csname url\endcsname\relax
  \def\url#1{\texttt{#1}}\fi
\expandafter\ifx\csname urlprefix\endcsname\relax\def\urlprefix{URL }\fi
\providecommand{\bibinfo}[2]{#2}
\providecommand{\eprint}[2][]{\url{#2}}

\bibitem[{\citenamefont{Lubensky}(1975)}]{quench-on}
\bibinfo{author}{\bibfnamefont{T.~C.} \bibnamefont{Lubensky}},
  \bibinfo{journal}{Physical Review B} \textbf{\bibinfo{volume}{11}},
  \bibinfo{pages}{3573} (\bibinfo{year}{1975}).

\bibitem[{\citenamefont{Mukherjee and Basu}(2020)}]{quench-on1}
\bibinfo{author}{\bibfnamefont{S.}~\bibnamefont{Mukherjee}} \bibnamefont{and}
  \bibinfo{author}{\bibfnamefont{A.}~\bibnamefont{Basu}},
  \bibinfo{journal}{Phys. Rev. Research} \textbf{\bibinfo{volume}{2}},
  \bibinfo{pages}{033423} (\bibinfo{year}{2020}).

\bibitem[{\citenamefont{Grinstein and Luther}(1976)}]{rand-field}
\bibinfo{author}{\bibfnamefont{G.}~\bibnamefont{Grinstein}} \bibnamefont{and}
  \bibinfo{author}{\bibfnamefont{A.}~\bibnamefont{Luther}},
  \bibinfo{journal}{Physical Review B} \textbf{\bibinfo{volume}{13}},
  \bibinfo{pages}{1329} (\bibinfo{year}{1976}).

\bibitem[{\citenamefont{Jeong et~al.}(1996)\citenamefont{Jeong, Kahng, and
  Kim}}]{kim96}
\bibinfo{author}{\bibfnamefont{H.}~\bibnamefont{Jeong}},
  \bibinfo{author}{\bibfnamefont{B.}~\bibnamefont{Kahng}}, \bibnamefont{and}
  \bibinfo{author}{\bibfnamefont{D.}~\bibnamefont{Kim}},
  \bibinfo{journal}{Physical review letters} \textbf{\bibinfo{volume}{77}},
  \bibinfo{pages}{5094} (\bibinfo{year}{1996}).

\bibitem[{\citenamefont{Csah{\'o}k et~al.}(1993)\citenamefont{Csah{\'o}k,
  Honda, Somfai, Vicsek, and Vicsek}}]{vicsek93}
\bibinfo{author}{\bibfnamefont{Z.}~\bibnamefont{Csah{\'o}k}},
  \bibinfo{author}{\bibfnamefont{K.}~\bibnamefont{Honda}},
  \bibinfo{author}{\bibfnamefont{E.}~\bibnamefont{Somfai}},
  \bibinfo{author}{\bibfnamefont{M.}~\bibnamefont{Vicsek}}, \bibnamefont{and}
  \bibinfo{author}{\bibfnamefont{T.}~\bibnamefont{Vicsek}},
  \bibinfo{journal}{Physica A: Statistical Mechanics and its Applications}
  \textbf{\bibinfo{volume}{200}}, \bibinfo{pages}{136} (\bibinfo{year}{1993}).

\bibitem[{\citenamefont{Barab{\'a}si et~al.}(1996)\citenamefont{Barab{\'a}si,
  Grinstein, and Munoz}}]{barabasi96}
\bibinfo{author}{\bibfnamefont{A.-L.} \bibnamefont{Barab{\'a}si}},
  \bibinfo{author}{\bibfnamefont{G.}~\bibnamefont{Grinstein}},
  \bibnamefont{and} \bibinfo{author}{\bibfnamefont{M.}~\bibnamefont{Munoz}},
  \bibinfo{journal}{Physical review letters} \textbf{\bibinfo{volume}{76}},
  \bibinfo{pages}{1481} (\bibinfo{year}{1996}).

\bibitem[{\citenamefont{Amaral et~al.}(1994)\citenamefont{Amaral, Barab{\'a}si,
  and Stanley}}]{barabasi94}
\bibinfo{author}{\bibfnamefont{L.~A.~N.} \bibnamefont{Amaral}},
  \bibinfo{author}{\bibfnamefont{A.-L.} \bibnamefont{Barab{\'a}si}},
  \bibnamefont{and} \bibinfo{author}{\bibfnamefont{H.~E.}
  \bibnamefont{Stanley}}, \bibinfo{journal}{Physical review letters}
  \textbf{\bibinfo{volume}{73}}, \bibinfo{pages}{62} (\bibinfo{year}{1994}).

\bibitem[{\citenamefont{Haldar and Basu}(2020)}]{astik-prr}
\bibinfo{author}{\bibfnamefont{A.}~\bibnamefont{Haldar}} \bibnamefont{and}
  \bibinfo{author}{\bibfnamefont{A.}~\bibnamefont{Basu}},
  \bibinfo{journal}{Phys. Rev. Research} \textbf{\bibinfo{volume}{2}},
  \bibinfo{pages}{043073} (\bibinfo{year}{2020}).

\bibitem[{\citenamefont{Tripathy and Barma}(1997)}]{mustansir1}
\bibinfo{author}{\bibfnamefont{G.}~\bibnamefont{Tripathy}} \bibnamefont{and}
  \bibinfo{author}{\bibfnamefont{M.}~\bibnamefont{Barma}},
  \bibinfo{journal}{Physical review letters} \textbf{\bibinfo{volume}{78}},
  \bibinfo{pages}{3039} (\bibinfo{year}{1997}).

\bibitem[{\citenamefont{Tripathy and Barma}(1998)}]{mustansir2}
\bibinfo{author}{\bibfnamefont{G.}~\bibnamefont{Tripathy}} \bibnamefont{and}
  \bibinfo{author}{\bibfnamefont{M.}~\bibnamefont{Barma}},
  \bibinfo{journal}{Physical Review E} \textbf{\bibinfo{volume}{58}},
  \bibinfo{pages}{1911} (\bibinfo{year}{1998}).

\bibitem[{\citenamefont{de~Queiroz and Stinchcombe}(2008)}]{stinchcombe}
\bibinfo{author}{\bibfnamefont{S.}~\bibnamefont{de~Queiroz}} \bibnamefont{and}
  \bibinfo{author}{\bibfnamefont{R.}~\bibnamefont{Stinchcombe}},
  \bibinfo{journal}{Physical Review E} \textbf{\bibinfo{volume}{78}},
  \bibinfo{pages}{031106} (\bibinfo{year}{2008}).

\bibitem[{\citenamefont{Kardar et~al.}(1986)\citenamefont{Kardar, Parisi, and
  Zhang}}]{kpz}
\bibinfo{author}{\bibfnamefont{M.}~\bibnamefont{Kardar}},
  \bibinfo{author}{\bibfnamefont{G.}~\bibnamefont{Parisi}}, \bibnamefont{and}
  \bibinfo{author}{\bibfnamefont{Y.-C.} \bibnamefont{Zhang}},
  \bibinfo{journal}{Physical Review Letters} \textbf{\bibinfo{volume}{56}},
  \bibinfo{pages}{889} (\bibinfo{year}{1986}).

\bibitem[{\citenamefont{Frey and T\"auber}(1994)}]{Frey-two-loop}
\bibinfo{author}{\bibfnamefont{E.}~\bibnamefont{Frey}} \bibnamefont{and}
  \bibinfo{author}{\bibfnamefont{U.~C.} \bibnamefont{T\"auber}},
  \bibinfo{journal}{Phys. Rev. E} \textbf{\bibinfo{volume}{50}},
  \bibinfo{pages}{1024} (\bibinfo{year}{1994}).

\bibitem[{\citenamefont{Chaikin and Lubensky}(2000)}]{chaikin}
\bibinfo{author}{\bibfnamefont{P.~M.} \bibnamefont{Chaikin}} \bibnamefont{and}
  \bibinfo{author}{\bibfnamefont{T.~C.} \bibnamefont{Lubensky}},
  \emph{\bibinfo{title}{Principles of condensed matter physics}},
  vol.~\bibinfo{volume}{1} (\bibinfo{publisher}{Cambridge university press
  Cambridge}, \bibinfo{year}{2000}).

\bibitem[{\citenamefont{Barab{\'a}si and Stanley}(1995)}]{stanley}
\bibinfo{author}{\bibfnamefont{A.-L.} \bibnamefont{Barab{\'a}si}}
  \bibnamefont{and} \bibinfo{author}{\bibfnamefont{H.~E.}
  \bibnamefont{Stanley}}, \emph{\bibinfo{title}{Fractal concepts in surface
  growth}} (\bibinfo{publisher}{Cambridge university press},
  \bibinfo{year}{1995}).

\bibitem[{\citenamefont{Forster et~al.}(1977)\citenamefont{Forster, Nelson, and
  Stephen}}]{forster}
\bibinfo{author}{\bibfnamefont{D.}~\bibnamefont{Forster}},
  \bibinfo{author}{\bibfnamefont{D.~R.} \bibnamefont{Nelson}},
  \bibnamefont{and} \bibinfo{author}{\bibfnamefont{M.~J.}
  \bibnamefont{Stephen}}, \bibinfo{journal}{Physical Review A}
  \textbf{\bibinfo{volume}{16}}, \bibinfo{pages}{732} (\bibinfo{year}{1977}).

\bibitem[{\citenamefont{L{\"a}ssig and Kinzelbach}(1997)}]{lassig97}
\bibinfo{author}{\bibfnamefont{M.}~\bibnamefont{L{\"a}ssig}} \bibnamefont{and}
  \bibinfo{author}{\bibfnamefont{H.}~\bibnamefont{Kinzelbach}},
  \bibinfo{journal}{Physical review letters} \textbf{\bibinfo{volume}{78}},
  \bibinfo{pages}{903} (\bibinfo{year}{1997}).

\bibitem[{\citenamefont{Bhattacharjee}(1998)}]{jkb98}
\bibinfo{author}{\bibfnamefont{J.}~\bibnamefont{Bhattacharjee}},
  \bibinfo{journal}{Journal of Physics A: Mathematical and General}
  \textbf{\bibinfo{volume}{31}}, \bibinfo{pages}{L93} (\bibinfo{year}{1998}).

\bibitem[{\citenamefont{Moser et~al.}(1991)\citenamefont{Moser, Kert{\'e}sz,
  and Wolf}}]{moser91}
\bibinfo{author}{\bibfnamefont{K.}~\bibnamefont{Moser}},
  \bibinfo{author}{\bibfnamefont{J.}~\bibnamefont{Kert{\'e}sz}},
  \bibnamefont{and} \bibinfo{author}{\bibfnamefont{D.~E.} \bibnamefont{Wolf}},
  \bibinfo{journal}{Physica A: Statistical Mechanics and its Applications}
  \textbf{\bibinfo{volume}{178}}, \bibinfo{pages}{215} (\bibinfo{year}{1991}).

\bibitem[{\citenamefont{Marinari et~al.}(2002)\citenamefont{Marinari, Pagnani,
  Parisi, and R{\'a}cz}}]{marinari02}
\bibinfo{author}{\bibfnamefont{E.}~\bibnamefont{Marinari}},
  \bibinfo{author}{\bibfnamefont{A.}~\bibnamefont{Pagnani}},
  \bibinfo{author}{\bibfnamefont{G.}~\bibnamefont{Parisi}}, \bibnamefont{and}
  \bibinfo{author}{\bibfnamefont{Z.}~\bibnamefont{R{\'a}cz}},
  \bibinfo{journal}{Physical review E} \textbf{\bibinfo{volume}{65}},
  \bibinfo{pages}{026136} (\bibinfo{year}{2002}).

\bibitem[{\citenamefont{Wiese}(1998)}]{wiese98}
\bibinfo{author}{\bibfnamefont{K.~J.} \bibnamefont{Wiese}},
  \bibinfo{journal}{Journal of statistical physics}
  \textbf{\bibinfo{volume}{93}}, \bibinfo{pages}{143} (\bibinfo{year}{1998}).

\bibitem[{\citenamefont{Zinn-Justin}(2002)}]{zinn}
\bibinfo{author}{\bibfnamefont{J.}~\bibnamefont{Zinn-Justin}},
  \emph{\bibinfo{title}{Quantum field theory and critical phenomena}}, vol.
  \bibinfo{volume}{113} (\bibinfo{publisher}{Clarendon Press, Oxford},
  \bibinfo{year}{2002}).

\bibitem[{\citenamefont{T{\"a}uber}(2014)}]{tauber}
\bibinfo{author}{\bibfnamefont{U.~C.} \bibnamefont{T{\"a}uber}},
  \emph{\bibinfo{title}{Critical dynamics: a field theory approach to
  equilibrium and non-equilibrium scaling behavior}}
  (\bibinfo{publisher}{Cambridge University Press}, \bibinfo{year}{2014}).

\bibitem[{\citenamefont{DeDominicis and Martin}(1979)}]{dedominicis}
\bibinfo{author}{\bibfnamefont{C.}~\bibnamefont{DeDominicis}} \bibnamefont{and}
  \bibinfo{author}{\bibfnamefont{P.}~\bibnamefont{Martin}},
  \bibinfo{journal}{Physical Review A} \textbf{\bibinfo{volume}{19}},
  \bibinfo{pages}{419} (\bibinfo{year}{1979}).

\bibitem[{\citenamefont{Bausch et~al.}(1976)\citenamefont{Bausch, Janssen, and
  Wagner}}]{bausch}
\bibinfo{author}{\bibfnamefont{R.}~\bibnamefont{Bausch}},
  \bibinfo{author}{\bibfnamefont{H.-K.} \bibnamefont{Janssen}},
  \bibnamefont{and} \bibinfo{author}{\bibfnamefont{H.}~\bibnamefont{Wagner}},
  \bibinfo{journal}{Zeitschrift f{\"u}r Physik B Condensed Matter}
  \textbf{\bibinfo{volume}{24}}, \bibinfo{pages}{113} (\bibinfo{year}{1976}).

\end{thebibliography}

\end{document}